\documentclass[aps,prb,reprint,superscriptaddress,nofootinbib,longbibliography]{revtex4-2}

\usepackage{amsmath,amssymb,bm}
\usepackage{graphicx}
\usepackage{microtype}
\usepackage{booktabs}
\usepackage[colorlinks=true,allcolors=blue]{hyperref}

\DeclareMathOperator{\sech}{sech}
\DeclareMathOperator{\sgn}{sgn}
\newcommand{\kb}{k_{\mathrm B}}
\newcommand{\bs}{B_{\mathrm s}}
\newcommand{\As}{\bm A_{\mathrm s}}
\newcommand{\DB}{D_B}
\newcommand{\cth}{\mathcal C_{\mathrm{th}}}
\newcommand{\ctr}{\mathcal C_{\mathrm{tr}}}
\newcommand{\sm}{s_m}
\newcommand{\Kg}{K_g}
\newcommand{\rhoM}{\rho_M}

\begin{document}

\title{Static and Ensemble-Dependent Thermodynamics of the Strain-Induced Parity Anomaly in Gapped Graphene}

\author{Ara~Sedrakyan}
\affiliation{Yerevan Physics Institute, Br.~Alikhanian 2, Yerevan 36, Armenia}

\author{Klaus~Ziegler}
\affiliation{Physics Department, New York City College of Technology, The City University of New York, Brooklyn, NY 11201, USA}
\affiliation{Institut f\"ur Physik, Universit\"at Augsburg, D-86135 Augsburg, Germany}

\date{\today}

\begin{abstract}
A static deformation of graphene can act on its Dirac electrons as a valley-odd magnetic field.  In sublattice-gapped graphene this field makes the two valleys add in the parity-odd response rather than cancel.  We derive the equilibrium thermodynamics of this effect and separate it from the finite-frequency transport response.  At fixed electrochemical potential, reversing the pseudomagnetic field removes every nonzero pseudo-Landau level in the continuum theory.  The remaining grand-potential difference is fixed by the spectrally asymmetric zeroth level.  The static charge response is a thermally broadened plateau confined to the gap and has no metallic $m/|\mu|$ tail.  Near a band edge, pseudofield reversal transfers $\kb\ln2$ of entropy per unsplit zero-mode state in the low-temperature window.  The fixed-$\mu$ heat capacity has two side lobes per edge and a universal peak $0.439229\,\DB\kb$.  We then formulate a definite constant-gate-voltage circuit and show that the measured sheet heat capacity depends on the electrical boundary condition.  The full massive-Dirac density of states and an exact finite-field pseudo-Landau-level calculation give the same gate crossover in their common limit.  At fixed carrier number, the low-temperature edge value is $-2(\ln2)^2\DB\kb$, rather than a node.  A finite geometric capacitance gives a continuous and experimentally tunable interpolation.  Finally, we give a trace-free triaxial strain geometry, a disorder--interaction window, and realistic calorimetric and quantum-capacitance scales.  The field-reversal protocol isolates an equilibrium electromechanical anomaly without a real magnetic field.
\end{abstract}

\maketitle

\section{Introduction}

Graphene and related two-dimensional materials have low-energy quasiparticles that obey a Dirac equation \cite{CastroNeto2009}.  A mass gap in one two-dimensional Dirac cone produces a parity-odd current response.  In field-theory language this is the parity anomaly and its Chern--Simons term \cite{NiemiSemenoff1983,Redlich1984PRL,Semenoff1984,Haldane1988}.  Sublattice gaps can arise when graphene is placed on an inversion-breaking substrate such as hexagonal boron nitride, and graphene--hBN devices now provide clean platforms for massive Dirac fermions \cite{Giovannetti2007,Hunt2013,Woods2014,Yankowitz2019}.  A real crystal contains both valleys and both spins, so the response of one cone is not by itself an observable charge response.  The signs carried by all species decide whether their contributions add or cancel.

Strain changes this sign rule.  Long-wavelength lattice deformation produces scalar, metric, and gauge couplings in graphene \cite{Manes2007,Vozmediano2010,deJuan2012,Amorim2016,SedrakyanSinnerZiegler2021,Sedrakyan2026FiniteFrequency}.  The gauge part has opposite sign in the two valleys and can generate pseudo-Landau levels without a real magnetic field \cite{PereiraCastroNeto2009,Guinea2010,GuineaGeim2010,Levy2010,Settnes2016,Berman-2022}.  For a sublattice, or Semenoff, mass, the reversal of valley chirality is accompanied by a reversal of gauge charge.  Their product is the same in both valleys.  A valley-odd mechanical field can therefore convert the one-cone parity anomaly into a valley-additive electrical response while preserving microscopic time-reversal symmetry.

The sublattice polarization of the zeroth Dirac Landau level is the microscopic source of this addition.  In a real magnetic field it also produces flux-quantized sublattice charge transfer.  Recent continuum and Hofstadter calculations showed one electron per spin transferred between sublattices when the flux changes by one quantum \cite{TarafdarSedrakyan2025SCO}.  The present work uses the axial, rather than real-field, version of the same spectral asymmetry and asks how it appears in equilibrium thermodynamics.

This problem follows a line of work by the authors and collaborators on deformation, geometry, and Dirac response.  Earlier results include the optical response to random lattice deformation \cite{SinnerSedrakyanZiegler2011}, the massive-Dirac current correlator at finite chemical potential \cite{ApresyanKhachatryanSedrakyan2015}, geometrically disordered and kagome quantum-Hall network models \cite{Gruzberg2017,Charles2020}, and the derivation of geometric electron--phonon gauge vertices in a deformed graphene sheet \cite{SedrakyanSinnerZiegler2021}.  The two papers that directly motivate the present work derived the finite-frequency anomaly-induced electromechanical response and its finite-temperature transport coefficient \cite{Sedrakyan2026FiniteFrequency,SedrakyanZiegler2026Thermal}.

Related work developed Chern--Simons descriptions of graphene and quantum magnets \cite{MaitiSedrakyan2019,SedrakyanMoessnerKamenev2020,WangXieWangSedrakyan2022}, and studied weak-field interaction, Friedel-oscillation, and ballistic-transport effects in graphene \cite{WangRaikhSedrakyan2021Friedel,WangRaikhSedrakyan2021Interaction,WangSedrakyan2022}.  These studies make clear that spectral chirality, gauge structure, interactions, and the order of infrared limits must be kept distinct.  The standard Dirac-Landau spectrum and its interaction scales are reviewed in Refs.~\cite{GusyninSharapov2005,Goerbig2011}, while intervalley scattering sets an important limit on any axial-field protocol \cite{MorpurgoGuinea2006}.

Strain engineering has continued to develop rapidly.  Recent work has provided a symmetry-based straintronics framework, studied nonlinear acoustic valley Hall response, and analyzed differential entropy in Dirac materials \cite{Zemouri2025,Wan2025,Chaika2025}.  Recent studies have examined periodic-strain flat bands, deformed graphene nanobubbles, and nonlinear-strain topological phases \cite{Sun2025JPCM,JungMyoung2025,Azizi2025}.  Quantum-capacitance measurements already resolve small changes in the graphene density of states \cite{Kretinin2013}.  These developments make a static thermodynamic test of a strain gauge field timely.

The question addressed here is simple but subtle: what does the anomaly do in equilibrium?  
The parity anomaly creates a jump in the Hall conductivity at $m=0$. Moreover, exchanging
the order of the limits $\bm q\to0$ and $\omega\to0$ gives different results:

A uniform transport coefficient is obtained by taking $\bm q\to0$ before $\omega\to0$.  A static thermodynamic response sets $\omega=0$ first and probes a slowly varying spatial field.  A doped Dirac metal is nonanalytic at $(\omega,\bm q)=(0,0)$, so these procedures need not agree \cite{Streda1982,Sharapov2015,XiaoChangNiu2010,QinNiuShi2011}.  At zero temperature they coincide inside the gap and differ in a band.  At finite temperature they are exponentially close only when the chemical potential lies many $\kb T$ inside the gap; they already separate near a thermally rounded band edge.

Our main results are as follows.  First, the pseudofield reversal gives an exact fixed-$\mu$ continuum expression for the field-odd grand potential.  Second, the static anomaly has no algebraic metallic tail.  Third, the zeroth pseudo-Landau level produces a flux-counting entropy and a sharp heat-capacity pattern.  Fourth, a definite gate circuit changes that pattern because the two field orientations acquire different chemical potentials.  We retain the full massive-Dirac density of states, solve the finite-field number constraint separately for the two orientations, and identify a controlled field--area window.  Finally, we give a trace-free triaxial deformation for which the pseudofield reverses while the leading scalar strain and elastic energy remain unchanged.

\section{Dirac model and valley selection rule}

We separate the one-particle spectrum from the thermodynamic chemical potential.  With the
valley  index $\tau=\pm1$ the one-particle Hamiltonian is
\begin{equation}
\begin{aligned}
h_\tau&=v_F\left(\tau\sigma_x\Pi_{\tau x}+\sigma_y\Pi_{\tau y}\right)
       +m\sigma_z,\\
\bm\Pi_\tau&=-i\hbar\bm\nabla+\tau e\As .
\end{aligned}
\label{eq:Hamiltonian}
\end{equation}
The many-body Hamiltonian is obtained by second quantization of $h_\tau$, and the grand-canonical operator is $\mathcal K=\mathcal H-\mu N$.  Thus $\mu$ enters only through occupations and thermodynamic derivatives; it is not included in the level energies.  Here $v_F$ is the 
 Fermi
velocity, $m$ is the sublattice mass, and $\mu$ is the electrochemical potential.  At $T=0$, we denote its value by the Fermi energy, $E_F\equiv\mu(T=0)$.  We use $n$ for electron number density, so the electrical charge density is $-en$.

The vector potential $\As$ is the equivalent strain gauge field and
\begin{equation}
\bs=(\bm\nabla\times\As)_z
\label{eq:Bsdef}
\end{equation}
is the corresponding equivalent pseudomagnetic field.  For the geometric phonon gauge field of Refs.~\cite{SedrakyanSinnerZiegler2021,Sedrakyan2026FiniteFrequency}, one may write $e\As=g_{\rm ph}\bm{\mathcal A}_{\rm ph}$.  The microscopic coupling $g_{\rm ph}$ is then absorbed into the experimentally calibrated field $\bs$.

It is useful to state the sign rule in a form that also applies to other Dirac materials.  Let a Dirac species $a$ have chirality $\eta_a=\pm1$, gauge-charge sign $\zeta_a=\pm1$, and mass $m_a$. Then the field-odd zeroth-level response is weighted by
\begin{equation}
\chi_a=\eta_a\zeta_a\sgn(m_a).
\label{eq:selection}
\end{equation}
Species with opposite $\chi_a$ cancel  upon summation.
In graphene with a real magnetic field, $\eta_\tau=\tau$ while $\zeta_\tau$ is valley independent.  A Semenoff mass then gives opposite zeroth-level shifts in the two valleys.  For the strain field in Eq.~\eqref{eq:Hamiltonian}, however, $\zeta_\tau=\tau$.  Thus $\eta_\tau\zeta_\tau=1$, and the valleys add.

For a uniform field, the pseudo-Landau levels, measured from the Dirac point before subtracting $\mu$, are
\begin{equation}
\begin{aligned}
E_{n,\pm}&=\pm\sqrt{M^2+2n\hbar e v_F^2|\bs|},\qquad n\geq1,\\
E_0&=-m\,\sgn(\bs),\qquad M\equiv |m|.
\end{aligned}
\label{eq:levels}
\end{equation}
The zeroth-level energy is the same in the two valleys, while the real spin gives a second degeneracy.  With $N_f=4$ spin-valley flavors, each level has degeneracy per area
\begin{equation}
\DB=\frac{N_f e|\bs|}{2\pi\hbar}
   =\frac{N_f|\bs|}{\Phi_0},
\qquad \Phi_0=\frac{h}{e}.
\label{eq:degeneracy}
\end{equation}
Each level with $n\geq1$ depends only on $|\bs|$.  At the same $T$ and $\mu$, these levels cancel exactly in the difference between opposite pseudofields.  Only the zeroth level changes energy.  Under a fixed-number or finite-gate constraint the two orientations generally have different chemical potentials, so the nonzero levels must then be retained; Sec.~\ref{sec:ensemble} treats this case exactly.

The continuum description requires strain that varies slowly on the lattice scale and weak intervalley scattering.  For a local-density treatment of a nonuniform field, the variation length should also exceed the magnetic length $\ell_s=\sqrt{\hbar/(e|\bs|)}$ and the mass correlation length $\xi_m=\hbar v_F/M$.  We discuss a quantitative strain profile in Sec.~\ref{sec:geometry}.

\section{Exact pseudofield-odd grand potential}

We define the full field-reversal difference
\begin{equation}
\Delta X=X(+|\bs|)-X(-|\bs|).
\label{eq:deltadef}
\end{equation}
For a level of energy $E$ and degeneracy $\DB$, the grand-potential density is
\begin{equation}
\omega_E=-\DB\kb T\ln\!\left[1+e^{-(E-\mu)/(\kb T)}\right].
\label{eq:onelevelomega}
\end{equation}
Let $\sm=\sgn(m)$ and
\begin{equation}
x_\pm=\frac{\mu\pm M}{2\kb T}.
\label{eq:xpm}
\end{equation}
Then the raw zeroth-level difference is
\begin{equation}
\begin{aligned}
\Delta\omega_{\rm raw}={}&-\DB\sm M\\
&-\DB\sm\kb T
 \left[\ln\cosh x_+-\ln\cosh x_-\right].
\end{aligned}
\label{eq:rawomega}
\end{equation}
The first term is independent of $T$ and $\mu$.  In a continuum Dirac theory its absolute value depends on the ultraviolet vacuum convention.  A gate-referenced experiment removes it by subtracting the value at charge neutrality.  We therefore use
\begin{equation}
\Delta\omega
=-\DB\sm\kb T
 \left[\ln\cosh x_+-\ln\cosh x_-\right].
\label{eq:freeenergy}
\end{equation}
All density, entropy, and heat-capacity results are independent of the discarded constant.  Equation~\eqref{eq:freeenergy} is exact for the part odd under pseudofield reversal in the uniform continuum spectrum and not just a weak-field expansion.

For a smooth field of one sign, the spatially integrated result is controlled by the number of zero modes.  The Aharonov--Casher index gives $N_0=N_f|\Phi_s|/\Phi_0$, up to boundary corrections, where $\Phi_s=\int d^2r\,\bs(\bm r)$ \cite{AharonovCasher1979}.  This gives a direct meaning to the factor $\DB A$ ($A$ is area): it is the number of states that take part in the field-reversal spectral flow. 

\section{Static thermodynamics versus uniform transport}
\label{sec:limits}

The equilibrium charge-density difference follows from Eq.~\eqref{eq:freeenergy}
and reads
\begin{equation}
\frac{\Delta n}{\DB}
\equiv\cth
=\frac{\sm}{2}\left(\tanh x_+-\tanh x_-\right).
\label{eq:Cth}
\end{equation}
At zero temperature we get
\begin{equation}
\cth(T=0)=
\begin{cases}
\sm,&|E_F|<M,\\
0,&|E_F|>M.
\end{cases}
\label{eq:Cthzero}
\end{equation}
Thus the static response is a plateau confined to the gap.
On the other hand,
the uniform transport coefficient derived from the parity-odd current correlator is  \cite{ApresyanKhachatryanSedrakyan2015,SedrakyanZiegler2026Thermal}
\begin{equation}
\ctr(m,\mu,T)
=m\int_M^\infty\frac{d\epsilon}{\epsilon^2}
\left[f(-\epsilon-\mu)-f(\epsilon-\mu)\right],
\label{eq:Ctr}
\end{equation}
where $f(E)=[e^{E/(\kb T)}+1]^{-1}$.  At $T=0$ we get
\begin{equation}
\ctr(T=0)=
\begin{cases}
\sm,&|E_F|<M,\\
m/|E_F|,&|E_F|>M.
\end{cases}
\label{eq:Ctrzero}
\end{equation}
such that the two coefficients agree inside the gap and differ in a band
at $T=0$. At finite temperature they remain exponentially close only when $M-|\mu|\gg\kb T$.  Near a thermally rounded edge they differ even on the nominal insulating side.  The zero-temperature transport tail $m/|E_F|$ comes from the occupied Berry curvature of the Fermi sea and its Fermi-surface completion.  It is not an equilibrium density created by a static pseudoflux.

The difference is an order-of-limits effect.  Uniform transport takes $\bm q\rightarrow0$ before $\omega\rightarrow0$.  Static thermodynamics takes $\omega\rightarrow0$ first and then lets the field vary slowly in space.  The St\v{r}eda-type relation for the odd equilibrium density is
\begin{equation}
\left(\frac{\partial n_{\rm odd}}{\partial\bs}\right)_{\mu,T}
=\frac{N_f}{2\Phi_0}\cth,
\qquad
n_{\rm odd}=\frac{n(+\bs)-n(-\bs)}{2}.
\label{eq:streda}
\end{equation}
In a metal, Eq.~\eqref{eq:streda} is not the full uniform Hall transport coefficient.

\begin{figure*}[t]
\includegraphics[width=0.98\textwidth]{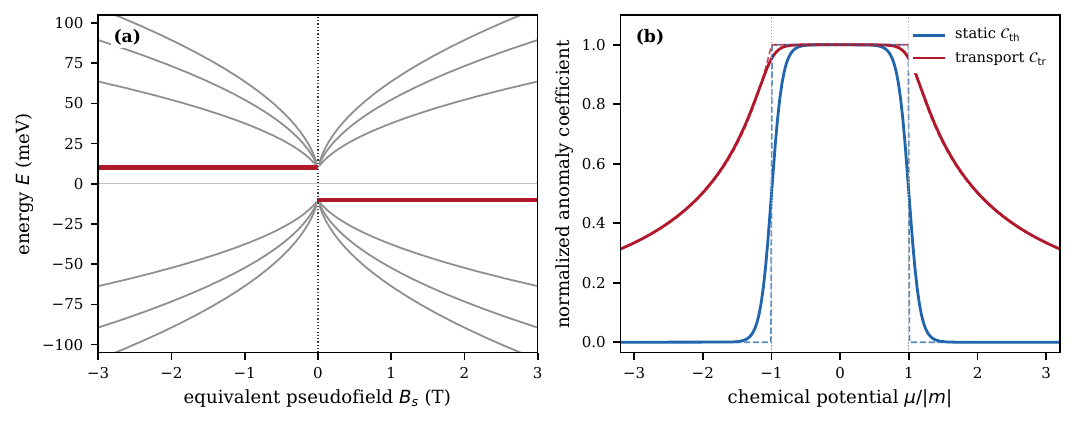}
\caption{\textbf{Spectral origin and noncommuting infrared limits.}
(a) The nonzero pseudo-Landau levels are even under $\bs\to-\bs$, while the zeroth level switches between $+m$ and $-m$.  The plot uses $M=10$ meV and $v_F=10^6$ m/s.  (b) The equilibrium coefficient $\cth$ and the uniform transport coefficient $\ctr$.  The dashed curves are the $T=0$ limits; the solid curves use $\kb T/M=0.08$.  The two responses coincide at $T=0$ inside the gap and are exponentially close deep in the finite-temperature gap.  Near an edge they separate before $|\mu|$ reaches $M$, and in a band only transport keeps the algebraic $m/|\mu|$ tail.}
\label{fig:limits}
\end{figure*}

Figure~\ref{fig:limits} shows both the spectral origin and the noncommuting limits.  This distinction is the first central result of the paper.  It also prevents a common mistake: one cannot obtain the equilibrium response by inserting a Fermi function into the zero-temperature transport coefficient.

\section{Entropy, heat capacity, and compressibility}
\label{sec:thermo}

All equilibrium observables follow by differentiating Eq.~\eqref{eq:freeenergy}.  
We define
\begin{equation}
g(x)=\ln\cosh x-x\tanh x,
\qquad
\phi(x)=x^2\sech^2x
\label{eq:kernels}
\end{equation}
and get
\begin{align}
\frac{\Delta s}{\DB\kb}
 &=\sm\left[g(x_+)-g(x_-)\right],
\label{eq:entropy}\\
\frac{\Delta c_\mu}{\DB\kb}
 &=\sm\left[\phi(x_+)-\phi(x_-)\right],
\label{eq:cmu}\\
\frac{\Delta\kappa}{\DB}
 &=\frac{\sm}{4\kb T}
 \left[\sech^2x_+-\sech^2x_-\right].
\label{eq:kappa}
\end{align}
Here $s$ is the entropy density, $c_\mu=T(\partial s/\partial T)_\mu$, and $\kappa=(\partial n/\partial\mu)_T$.  The thermal density response is
\begin{equation}
\begin{aligned}
\Delta\alpha
&\equiv\left(\frac{\partial\Delta n}{\partial T}\right)_\mu\\
&=\frac{\DB\sm}{2T}
\left[-x_+\sech^2x_++x_-\sech^2x_-\right].
\end{aligned}
\label{eq:alpha}
\end{equation}
These functions obey the Maxwell relation
\begin{equation}
\left(\frac{\partial\Delta s}{\partial\mu}\right)_T
=
\left(\frac{\partial\Delta n}{\partial T}\right)_\mu.
\label{eq:maxwell}
\end{equation}
At the conduction-band edge $\mu=M$, one zeroth level is at the electrochemical potential while its field-reversed partner is separated by almost the full gap $2M$.  For $\kb T\ll M$
we obtain
\begin{equation}
\Delta s(\mu=M)
=-\sm\DB\kb\ln2+O\!\left(e^{-2M/(\kb T)}\right).
\label{eq:entropyedge}
\end{equation}
The valence edge has the opposite sign.  Each state exactly at the electrochemical potential has occupation probability $1/2$ and entropy $\kb\ln2$.  The reversal difference therefore counts the zero-mode degeneracy.  This is a spectral, rather than an interaction, entropy.

The fixed-$\mu$ heat capacity has a different structure.  For a level exactly at $\mu$ has an entropy but its occupation does not change to first order with temperature, so its heat capacity vanishes.  Near the conduction edge we introduce
\begin{equation}
y=\frac{\mu-M}{2\kb T},
\qquad \kb T\ll M.
\label{eq:ydef}
\end{equation}
Then the universal edge functions are
\begin{align}
\frac{\Delta s}{\DB\kb\sm}
 &=-\ln2-g(y),
\label{eq:edges}\\
\frac{\Delta c_\mu}{\DB\kb\sm}
 &=-\phi(y),
\label{eq:edgec}\\
\frac{4\kb T\Delta\kappa}{\DB\sm}
 &=-\sech^2y.
\label{eq:edgek}
\end{align}
The valence edge follows by particle-hole transformation and  reversed sign.

\begin{figure*}[t]
\includegraphics[width=0.98\textwidth]{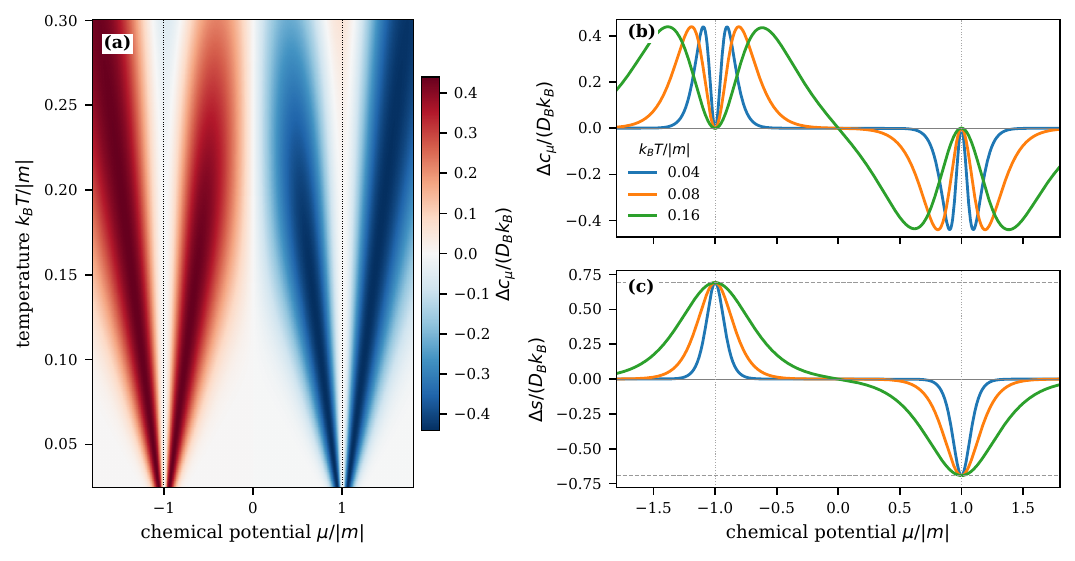}
\caption{\textbf{Thermodynamic response at fixed electrochemical potential.}
(a) Differential heat capacity as a function of chemical potential and temperature.  (b) At low temperature, two side lobes form around each band edge.  The apparent nodes at $\mu=\pm M$ become exact only in the scaling limit $\kb T/M\to0$; the finite-gap remnant is exponentially small.  (c) The entropy is concentrated near the two edges and approaches $\pm\DB\kb\ln2$.  The plotted quantities are exact for the part odd under pseudofield reversal in the uniform continuum model.}
\label{fig:thermo}
\end{figure*}

The extrema of $\phi(y)$ satisfy
\begin{equation}
y_*\tanh y_*=1,
\qquad y_*=1.1996786403,
\label{eq:ystar}
\end{equation}
and
\begin{equation}
\phi(y_*)=0.4392288399.
\label{eq:phistar}
\end{equation}
Thus each band edge has two fixed-$\mu$ side lobes at
\begin{equation}
|\mu\mp M|=2y_*\kb T
=2.39935728\,\kb T.
\label{eq:mupeak}
\end{equation}
Across both edges this gives the four-lobe pattern in Fig.~\ref{fig:thermo}.

At finite $T/M$, the remote zeroth level gives a small correction to the apparent node.  At 
$\mu=M$ we obtain
\begin{equation}
\frac{\Delta c_\mu}{\DB\kb\sm}
=\phi\!\left(\frac{M}{\kb T}\right)
\simeq
4\left(\frac{M}{\kb T}\right)^2e^{-2M/(\kb T)}.
\label{eq:finitegapnode}
\end{equation}
The original low-temperature statement should therefore be read as an asymptotic node, not an exact finite-temperature identity.  This correction is numerically negligible once $M/(\kb T)$ is several units  large, but it matters for a precise statement of the result.

The compressibility gives an independent and often easier signal.  Near the conduction edge it has a negative peak of magnitude $\DB/(4\kb T)$; the valence edge has the opposite sign.  Multiplying by $e^2$ gives the differential quantum capacitance per area.

\section{Gate circuit and electrical-ensemble crossover}
\label{sec:ensemble}

A calorimeter measures $c_\mu$ only when electrical contacts pin the graphene electrochemical potential.  A nearly isolated sheet is closer to fixed carrier number.  A common experiment lies between these limits: graphene is capacitively coupled to a gate held at fixed voltage.  We now formulate that circuit explicitly.

Let $C_g$ be the geometric capacitance per area and define the circuit charge susceptibility
\begin{equation}
\Kg=\frac{C_g}{e^2}.
\label{eq:Kg}
\end{equation}
For the pseudofield orientation $\sigma=\pm$, electrochemical equilibrium at fixed gate voltage is
\begin{equation}
\mu_\sigma+\frac{n_\sigma(T,\mu_\sigma)-n_g}{\Kg}=\mu_{\rm res}.
\label{eq:gateconstraint}
\end{equation}
Here $n_g$ is the density set by the gate offset and $\mu_{\rm res}$ is fixed by the voltage source.  We assume that the gate is an electrical reservoir but is not thermally modulated with the graphene sheet.  Its own heat capacity is then a field-even background.  If the gate is thermally coupled to the sheet, its ordinary heat capacity must be added to both orientations before taking their difference.

Differentiating Eq.~\eqref{eq:gateconstraint} at fixed gate voltage gives
\begin{equation}
\frac{d\mu_\sigma}{dT}
=-\frac{\alpha_\sigma}{\kappa_\sigma+\Kg},
\label{eq:dmudT}
\end{equation}
where $\alpha=(\partial n/\partial T)_\mu$ and $\kappa=(\partial n/\partial\mu)_T$.  The entropy derivative of the graphene sheet along this circuit path is therefore
\begin{equation}
c_{g,\sigma}=c_{\mu,\sigma}
-\frac{T\alpha_\sigma^2}{\kappa_\sigma+\Kg}.
\label{eq:cR}
\end{equation}
The limits $\Kg\to\infty$ and $\Kg\to0$ give fixed $\mu$ and fixed sheet density, respectively.  Equation~\eqref{eq:cR} is the sheet heat capacity measured in a constant-voltage protocol; it is not the heat capacity of a thermally equilibrated macroscopic reservoir.

For a weak local pseudofield, {\color{red} we} 
expand around the zero-field chemical potential $\mu_0$.  Let $D=\kappa_0+\Kg$, where the subscript $0$ denotes the field-even massive-Dirac background.  Keeping every term linear in the field-reversal difference gives
\begin{equation}
\begin{aligned}
\Delta c_g={}&\Delta c_\mu
-\frac{2T\alpha_0\Delta\alpha}{D}
+\frac{T\alpha_0^2\Delta\kappa}{D^2}\\
&-\frac{\Delta n}{D}\,
\partial_\mu\!\left(c_{\mu0}-\frac{T\alpha_0^2}{D}\right).
\end{aligned}
\label{eq:deltaCRgeneral}
\end{equation}
The last line is essential.  It is the change of the field-even heat capacity caused by the orientation-dependent chemical-potential shift
$\mu_+-\mu_-=-\Delta n/D$.  At a fixed $\mu$ this shift vanishes,  but at a fixed density it is of the same order as the direct odd terms.

We evaluate Eq.~\eqref{eq:deltaCRgeneral} with the full massive-Dirac density of states
\begin{equation}
\rho(E)=\rhoM\frac{|E|}{M}\,\Theta(|E|-M),
\qquad
\rhoM=\frac{N_fM}{2\pi\hbar^2v_F^2}.
\label{eq:rho0}
\end{equation}
The corresponding density, compressibility, thermal-density response, and heat capacity are given as convergent one-dimensional integrals in Appendix~\ref{app:ensemble}.  The gate strength is
\begin{equation}
\gamma=\frac{\Kg}{\rhoM}
=\frac{C_g}{e^2\rhoM}.
\label{eq:gammalambda}
\end{equation}
For $M=10$ meV and a $20$ nm hBN dielectric with relative permittivity $3.5$, $\gamma=0.658$.

The low-temperature limit is transparent.  Near the conduction edge we 
define $y=(\mu_0-M)/(2\kb T)$.  Then at fixed $\mu$ we get
\begin{equation}
\frac{\Delta c_\mu}{\DB\kb\sm}
=-y^2\sech^2y.
\label{eq:fixedmu_edge_again}
\end{equation}
A fixed carrier number and retaining the chemical-potential displacement gives
\begin{equation}
\frac{\Delta c_N}{\DB\kb\sm}
=-e^{-2y}\!
\left(1+e^{-2y}\right)
\ln^2\!\left(1+e^{2y}\right).
\label{eq:fixedn}
\end{equation}
Thus the fixed-$\mu$ edge node is removed.  At $y=0$ we have
\begin{equation}
\frac{\Delta c_N}{\DB\kb\sm}
=-2(\ln2)^2=-0.9609060278.
\label{eq:fixednedge}
\end{equation}
The factor of two compared with a calculation that keeps only the direct odd heat capacity comes from the shift of the field-even background in Eq.~\eqref{eq:deltaCRgeneral}.  The constant-density-of-states result is the $\kb T/M\to0$ scaling limit.  At $M=10$ meV and $T=10$ K, the full density of states changes the edge value to $-1.106166\,\DB\kb\sm$.

We also validate the expansion without using a density-of-states approximation.  For each field orientation we sum the full pseudo-Landau spectrum and solve Eq.~\eqref{eq:gateconstraint} separately.  It is useful to define
\begin{equation}
\epsilon_B=\frac{\hbar e v_F^2|\bs|}{M\kb T}.
\label{eq:epsilonB}
\end{equation}
Then at fixed density, $M=10$ meV, $T=10$ K, and $\epsilon_B=0.10$, the exact 
solution gives
\begin{equation}
\frac{\mu_+}{M}=0.9959803,
\qquad
\frac{\mu_-}{M}=1.0036929,
\label{eq:exactmus}
\end{equation}
and
\begin{equation}
\frac{\Delta c_N^{\rm exact}}{\DB\kb\sm}
=-1.1061547.
\label{eq:exactedge}
\end{equation}
The full-DOS field-linear value differs by only $1.2\times10^{-5}$, and the dimensionless number-constraint residual is below $3\times10^{-16}$.  Nonzero pseudo-Landau levels are included in this check; they no longer cancel term by term because $\mu_+\neq\mu_-$.

\begin{figure*}[t]
\includegraphics[width=0.98\textwidth]{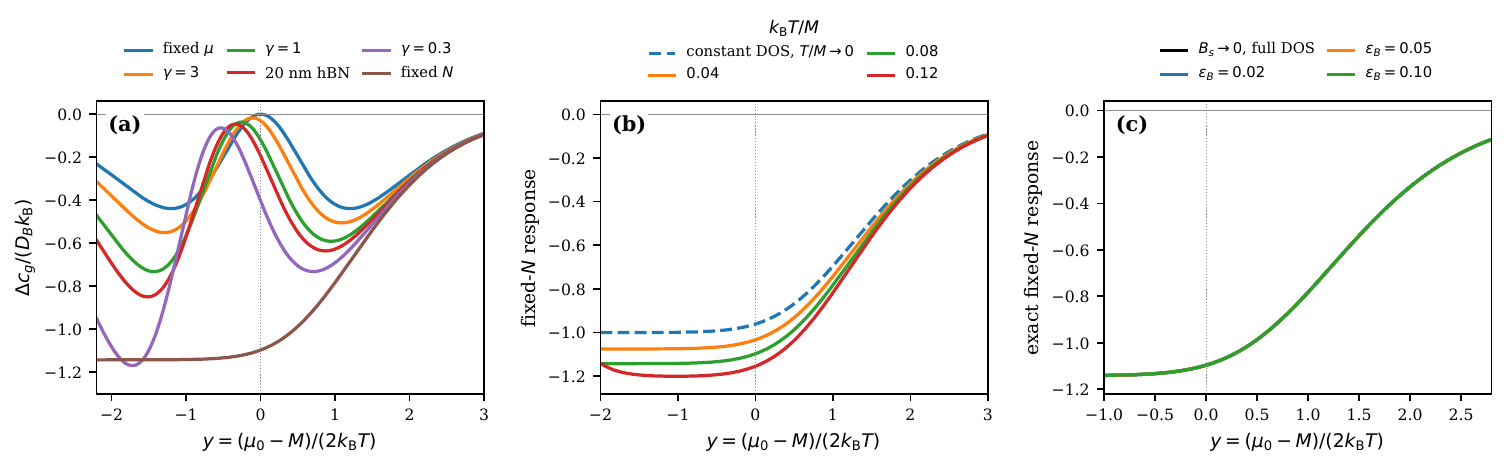}
\caption{\textbf{Gate circuit and exact ensemble crossover near the conduction edge.}
(a) Full massive-Dirac-density-of-states result at $\kb T/M=0.08617$.  The limits are fixed $\mu$ ($\gamma\to\infty$) and fixed carrier number ($\gamma=0$); a $20$ nm hBN gate gives $\gamma=0.658$.  (b) Fixed-number curves for several temperatures compared with the constant-density-of-states limit in Eq.~\eqref{eq:fixedn}.  (c) Exact finite-field pseudo-Landau-level sums at fixed number.  Solving the number constraint separately for the two orientations converges rapidly to the full-DOS $\bs\to0$ result.}
\label{fig:ensemble}
\end{figure*}

The field-linear expansion is controlled when the orientation-dependent chemical-potential shift is small compared with $\kb T$:
\begin{equation}
\epsilon_R
=\frac{\DB}{\kb T(\kappa_0+\Kg)}\ll1.
\label{eq:weakflux}
\end{equation}
Near an occupied edge and at fixed number, this is of the same order as $\epsilon_B$.  This is a condition on the local field, not on the total pseudoflux.  The signal still scales with 
$|\bs|A$, so one may reduce the field and increase the active area.  The exact pseudo-Landau calculation provides a direct check when $\epsilon_R$ is small but nonzero.  The fixed-$\mu$ thermodynamics of Sec.~\ref{sec:thermo} does not require this expansion because both orientations are compared at the same $\mu$.

\section{Disorder, interactions, and lattice corrections}
\label{sec:robustness}

\begin{figure}[t]
\includegraphics[width=\columnwidth]{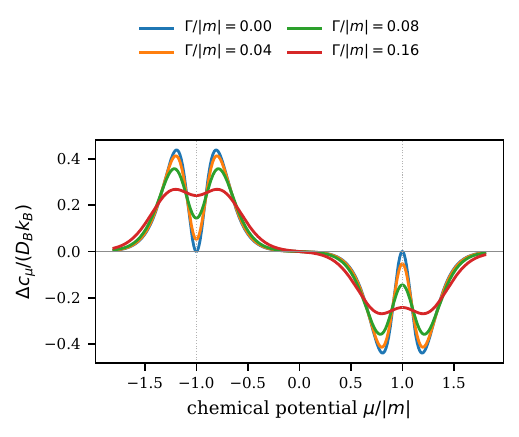}
\caption{Gaussian broadening of the fixed-$\mu$ differential heat capacity at $\kb T/M=0.08$.  Broadening rounds and merges the side lobes.  The clean response is recovered continuously as $\Gamma\to0$.}
\label{fig:disorder}
\end{figure}

A finite linewidth replaces the delta-function zeroth-level spectral weight by a normalized function.  For Gaussian broadening,
\begin{equation}
A_\Gamma(\delta)
=\frac{1}{\sqrt{2\pi}\Gamma}
 \exp\left(-\frac{\delta^2}{2\Gamma^2}\right),
\label{eq:gaussian}
\end{equation}
the correctly ordered fixed-$\mu$ convolution is
\begin{equation}
\begin{aligned}
\Delta c_\mu^{(\Gamma)}
=&\DB\kb\sm\int d\delta\,A_\Gamma(\delta)\\
&\times\left[
\phi\!\left(x_+-\frac{\delta}{2\kb T}\right)
-
\phi\!\left(x_--\frac{\delta}{2\kb T}\right)
\right].
\end{aligned}
\label{eq:broadening}
\end{equation}
This expression reduces to Eq.~\eqref{eq:cmu} with the correct sign as $\Gamma\to0$.  Figure~\ref{fig:disorder} shows that the lobe pattern remains clear for $\Gamma\lesssim\kb T$ and becomes smoother when the linewidth is larger.  The same spectral convolution can be inserted into the gate calculation before solving Eq.~\eqref{eq:gateconstraint}.

Interactions can split the highly degenerate zeroth pseudo-Landau level.  The clean single-particle line shapes describe the window
\begin{equation}
\Gamma,\ \Delta_{\rm int}\lesssim\kb T\ll M,
\label{eq:window}
\end{equation}
together with the circuit condition in Eq.~\eqref{eq:weakflux}.  In this window the nominal $\kb\ln2$ entropy is a finite-temperature crossover of an unresolved level.  It is not a protected residual entropy as $T\to0$: any resolved disorder or interaction splitting removes it.  If the splitting is larger than $\kb T$, the anomaly survives as additional many-body peaks rather than as the simple universal curve.  Interaction-driven spin and valley order in graphene Landau levels is well established \cite{Goerbig2011,RoyHuYang2013}.

A useful upper estimate is the Coulomb scale
\begin{equation}
E_C=\frac{e^2}{4\pi\epsilon_0\epsilon_r\ell_s}.
\label{eq:EC}
\end{equation}
A metallic gate with a distance $d_g$ away reduces long-wavelength exchange.  A simple image-charge estimate is
\begin{equation}
E_{\rm int}^{\rm scr}\simeq E_C\left(1-e^{-2d_g/\ell_s}\right).
\label{eq:screenedE}
\end{equation}
This expression is only a scale estimate, not a microscopic Hartree--Fock result.  It is conservative enough to identify a regime where thermal broadening dominates the interaction splitting.

The exact cancellation of all $n\geq1$ levels is a fixed-$\mu$ statement about the uniform continuum Hamiltonian.  A lattice model adds intervalley scattering, trigonal warping, boundaries, scalar deformation potentials, and nonlinear strain terms.  Intervalley scattering is especially harmful because it mixes the two species whose axial charges protect the additive response \cite{MorpurgoGuinea2006}.  The best protocol uses two mirror-related deformations with equal $|\bs|$ and equal trace strain.  Field-even elastic and electronic backgrounds then cancel to leading order.  Tight-binding calculations of triaxial pseudomagnetic dots support the continuum sublattice and valley structure \cite{Settnes2016}, while the recent Hofstadter calculation of sublattice charge order shows explicitly how zeroth-level spectral flow survives lattice regularization \cite{TarafdarSedrakyan2025SCO}.

\section{Quantitative strain geometry and geometric phonon flux}
\label{sec:geometry}

A differential thermodynamic experiment needs two deformations that reverse $\bs$ without changing large scalar and elastic backgrounds.  A trace-free triaxial in-plane displacement provides a simple design \cite{Guinea2010,Settnes2016}:
\begin{equation}
u_x=2Cxy,
\qquad
u_y=C(x^2-y^2).
\label{eq:triaxialu}
\end{equation}
Its linear strain tensor obeys
\begin{equation}
\begin{aligned}
u_{xx}&=2Cy, & u_{yy}&=-2Cy,\\
u_{xy}&=2Cx, & \mathrm{Tr}\,u&=0.
\end{aligned}
\label{eq:triaxialstrain}
\end{equation}
Using the standard graphene gauge coupling
$e\As=(\hbar\beta/2a)(u_{xx}-u_{yy},-2u_{xy})$, one obtains a uniform field
\begin{equation}
\bs=-\frac{4\hbar\beta C}{ea}.
\label{eq:triaxialB}
\end{equation}
Here $a=0.142$ nm is the carbon--carbon bond length and $\beta\simeq3$ is the hopping Gr\"uneisen parameter.  The largest principal strain in a circular membrane of radius $R$ is $\epsilon_{\max}=2|C|R$, so
\begin{equation}
|\bs|=\frac{2\hbar\beta\epsilon_{\max}}{eaR}.
\label{eq:triaxialBstrain}
\end{equation}
Changing $C\to-C$ reverses the pseudofield while leaving $\mathrm{Tr}\,u=0$, $u_{ij}u_{ij}$, and the leading elastic energy unchanged.  This directly realizes the mirror-reversal subtraction assumed above.

For the controlled fixed-number point used below, $|\bs|=1.309$ mT and $|\bs|A=100\ {\rm T}\,\mu{\rm m}^2$.  A circular active region then has
\begin{equation}
\begin{aligned}
A&=7.638\times10^4\,\mu{\rm m}^2,
&R&=155.9\,\mu{\rm m},\\
\epsilon_{\max}&=0.734\%,
&|u|_{\max}&=0.572\,\mu{\rm m}.
\end{aligned}
\label{eq:triaxialnumbers}
\end{equation}
The magnetic length is $\ell_s=709$ nm and the mass correlation length for $M=10$ meV is $\xi_m=65.8$ nm.  Both are much smaller than $R$, so the central region is locally uniform.  Nonlinear elastic corrections scale as $\epsilon_{\max}^2\simeq5.4\times10^{-5}$.

\begin{figure*}[t]
\includegraphics[width=0.98\textwidth]{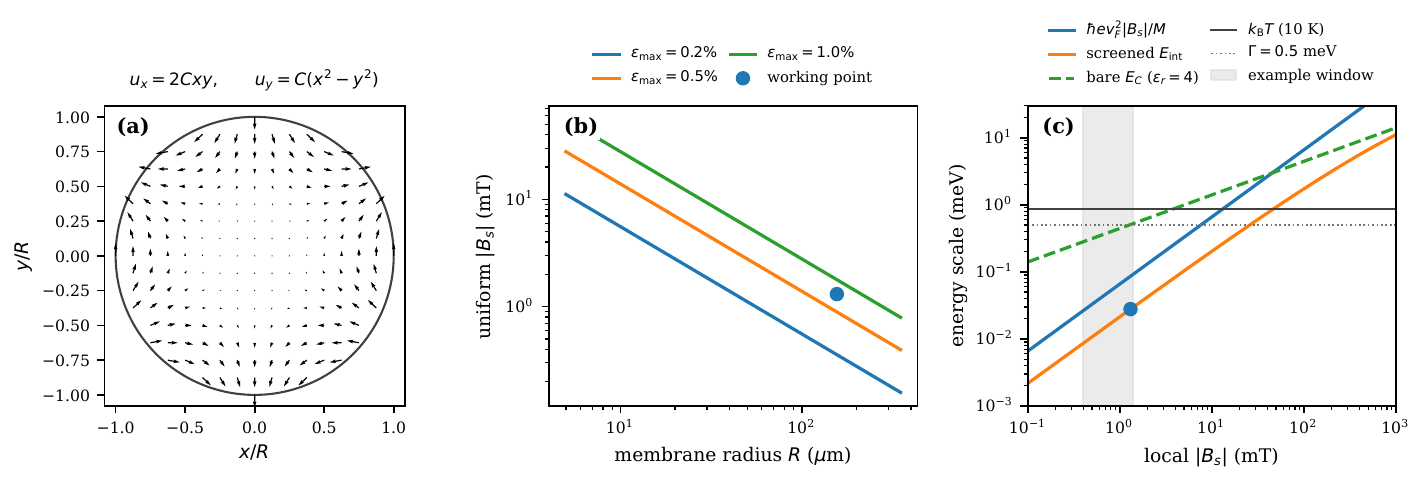}
\caption{\textbf{Trace-free triaxial design and feasibility window.}
(a) Displacement field $u_x=2Cxy$, $u_y=C(x^2-y^2)$ in a circular membrane.  Reversing $C$ reverses the uniform pseudofield while preserving the leading scalar strain and elastic energy.  (b) Uniform $|\bs|$ versus membrane radius for several maximum principal strains.  The dot marks $|\bs|=1.309$ mT, $R=155.9\,\mu$m, and $\epsilon_{\max}=0.734\%$.  (c) Cyclotron-edge, screened-interaction, bare Coulomb, thermal, and linewidth scales.  The shaded region contains the representative working point.}
\label{fig:triaxial}
\end{figure*}

For the geometric flexural gauge vertex derived in Refs.~\cite{SedrakyanSinnerZiegler2021,Sedrakyan2026FiniteFrequency}, the small-slope form is
\begin{equation}
\begin{aligned}
\mathcal A_i&=-\frac12(\partial_i h)\nabla^2h,\\
\mathcal B_{\mathcal A}
&=\epsilon_{ij}\partial_i\mathcal A_j
=\frac12\epsilon_{ij}(\partial_i h)\partial_j\nabla^2h.
\end{aligned}
\label{eq:geometricA}
\end{equation}
For two noncollinear flexural modes,
\begin{equation}
h(\bm r)=h_1\cos\theta_1+h_2\cos\theta_2,
\qquad
\theta_j=\bm q_j\cdot\bm r+\varphi_j,
\label{eq:twomodes}
\end{equation}
the cross term is
\begin{equation}
\begin{aligned}
\mathcal B_{\mathcal A}^{(12)}
={}&\frac{h_1h_2}{2}(q_1^2-q_2^2)\\
&\times(\bm q_1\times\bm q_2)_z
\sin\theta_1\sin\theta_2,
\end{aligned}
\label{eq:twoflux}
\end{equation}
 which requires non-collinear wave vectors with unequal magnitudes.  Restoring the microscopic coupling gives $\bs=(g_{\rm ph}/e)\mathcal B_{\mathcal A}$.

\begin{figure}[t]
\includegraphics[width=\columnwidth]{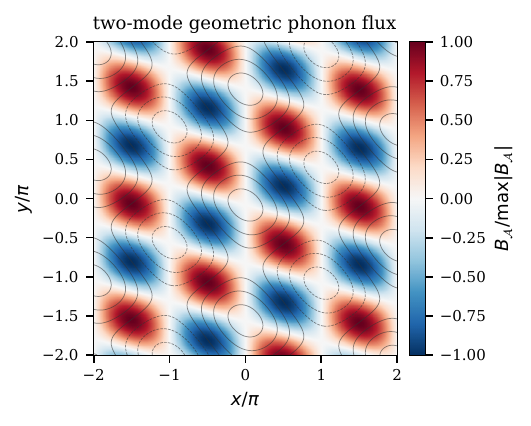}
\caption{Normalized geometric phonon flux from two non-collinear flexural modes.  Gray contours show the height field.  The alternating sign explains why a simple periodic texture needs a local probe, a finite envelope, or spatially weighted detection.}
\label{fig:geometry}
\end{figure}

An infinite two-mode texture normally has equal positive and negative flux.  It is therefore best studied by local heat-capacity or quantum-capacitance imaging, or by a detector whose spatial weight selects one Fourier component.  Triaxial strain is better suited to an integrated calorimeter because it creates a sign-definite central field.  Bent ribbons, elliptic nanobubbles, and periodic strain patterns provide complementary geometries \cite{GuineaGeim2010,JungMyoung2025,Sun2025JPCM}.

\section{Experimental scales and protocols}
\label{sec:experiment}

For graphene with $N_f=4$ we have
\begin{equation}
\begin{aligned}
\DB(1\ {\rm T})
&=9.67195697\times10^{14}\ {\rm m}^{-2}\\
&=9.67195697\times10^{10}\ {\rm cm}^{-2}
\end{aligned}
\label{eq:DBnumber}
\end{equation}
and for $|\bs|A=100\ {\rm T}\,\mu{\rm m}^2$, the spin-valley zero-mode count is
\begin{equation}
\DB A=9.67196\times10^4.
\label{eq:N0number}
\end{equation}
The low-temperature entropy difference at an edge is therefore
\begin{equation}
|\Delta S|_{\rm edge}
=\DB A\kb\ln2
=9.26\times10^{-19}\ {\rm J/K},
\label{eq:Snumber}
\end{equation}
provided any zero-level splitting remains below $\kb T$.

At fixed electrochemical potential, the peak heat-capacity difference is
\begin{equation}
|\Delta C_\mu|_{\max}
=0.43922884\,\DB A\kb
=5.87\times10^{-19}\ {\rm J/K}.
\label{eq:Cmunumber}
\end{equation}
For $M=10$ meV and $T=10$ K the extrema lie {\color{red} at} $2.07$ meV from each band edge.  The intrinsic differential quantum-capacitance peak is
\begin{equation}
e^2|\Delta\kappa|_{\max}
=4.50\,\mu{\rm F}/{\rm cm}^2
\label{eq:Cqnumber}
\end{equation}
for $|\bs|=1$ T and $T=10$ K.  A two-terminal capacitance experiment measures the series combination
\begin{equation}
C_{\rm tot}^{-1}=C_g^{-1}+C_Q^{-1},
\label{eq:seriescap}
\end{equation}
so a thin or high-permittivity dielectric, or a local charge sensor, improves the electrical readout \cite{Kretinin2013}.

The fixed-number experiment must use a weak local field.  The triaxial working point in Eq.~\eqref{eq:triaxialnumbers} has $\epsilon_B=0.10$ but the same total pseudoflux as $1$ T over $100\,\mu{\rm m}^2$.  At $M=10$ meV and $T=10$ K, the full-DOS edge signal is
\begin{equation}
|\Delta C_N|_{\rm edge}
=1.106166\,\DB A\kb
=1.48\times10^{-18}\ {\rm J/K}.
\label{eq:Cnnumber}
\end{equation}
The exact finite-field value differs by less than $2\times10^{-23}$ J/K for this flux-area product.  A $20$ nm hBN gate gives the intermediate edge signal
\begin{equation}
|\Delta C_g|_{\rm edge}
=0.198608\,\DB A\kb
=2.65\times10^{-19}\ {\rm J/K}.
\label{eq:Cgnumber}
\end{equation}
Changing the dielectric thickness therefore provides a direct circuit control of the line shape and amplitude.

At the triaxial working point, the bare Coulomb scale for $\epsilon_r=4$ is $0.508$ meV.  A metallic gate $20$ nm away reduces the image-charge estimate to $0.0278$ meV.  Both are below $\kb T=0.862$ meV after screening.  A representative linewidth $\Gamma=0.5$ meV also satisfies Eq.~\eqref{eq:window}.  The cyclotron-edge scale $\hbar e v_F^2|\bs|/M=0.0862$ meV is smaller than $\kb T$; resolved pseudo-Landau peaks are not required for the field-linear thermodynamic anomaly.  The hierarchy is instead designed to keep the orientation-dependent chemical-potential shift perturbative and the zero-level splitting thermally unresolved.

\begin{figure*}[t]
\includegraphics[width=0.98\textwidth]{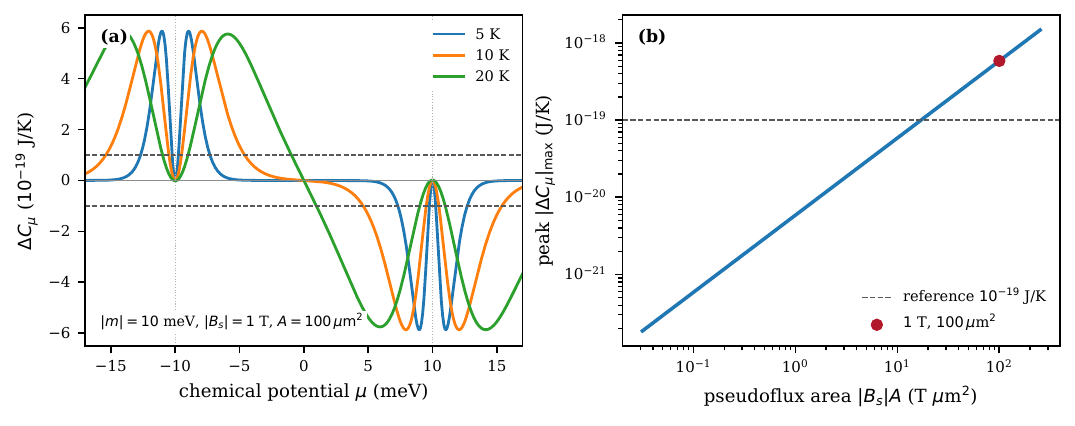}
\caption{\textbf{Calorimetric scale at fixed electrochemical potential.}
(a) Full heat-capacity reversal difference for $M=10$ meV, $|\bs|=1$ T, and $A=100\,\mu{\rm m}^2$.  Horizontal dashed lines mark $10^{-19}$ J/K.  (b) The universal peak grows linearly with the pseudofield-area product $|\bs|A$.  The fixed-number and finite-gate signals have the same total-flux scaling, but their line shapes also depend on the local-field and circuit parameters shown in Figs.~\ref{fig:ensemble} and \ref{fig:triaxial}.}
\label{fig:feasibility}
\end{figure*}

Graphene electronic calorimetry has reached heat-capacity sensitivity near and below $10^{-19}$ J/K \cite{FongSchwab2012,Aamir2021}.  Pseudomagnetic fields from tens to hundreds of tesla have been seen in small strained regions \cite{Levy2010,Nigge2019}; the present fixed-number protocol deliberately uses a much weaker field over a larger area.  The large membrane raises the ordinary lattice heat capacity, but that background is even under $C\to-C$.  The practical requirement is stable differential modulation: the two triaxial states must reproduce $|\bs|$, temperature, and contact conditions to better than the desired fractional anomaly signal.

A measurement can proceed in four steps.  First, open a clean sublattice gap using aligned hBN or another inversion-breaking substrate \cite{Giovannetti2007,Hunt2013,Woods2014}.  Second, switch between the two triaxial states $C$ and $-C$.  Their trace strain vanishes to linear order, while the pseudofield changes sign.  Third, sweep the gate through both band edges while recording heat capacity and capacitance.  Fourth, repeat with different $C_g$ or with the sheet electrically isolated.  A well-contacted device shows the fixed-$\mu$ four-lobe trace.  Reducing $C_g$ moves the response toward the fixed-number curve in Fig.~\ref{fig:ensemble}.  Reversing the mass sign or the strain chirality reverses every anomaly signal, while leading elastic and phonon backgrounds remain even.

\section{Conclusions}

We have derived the static thermodynamics of the strain-induced parity anomaly in gapped graphene.  At fixed electrochemical potential, pseudofield reversal removes every nonzero pseudo-Landau level in the uniform continuum model.  The remaining grand potential is fixed by the spectrally asymmetric zeroth level.

The equilibrium and transport anomalies are not the same in a doped cone.  At $T=0$ the static charge response ends at the band edge, while uniform transport keeps the Berry-curvature tail $m/|E_F|$.  At finite temperature the two limits also separate within a few $\kb T$ of an edge.  Thermodynamic derivatives produce a low-temperature $\kb\ln2$ entropy per unresolved zero-mode state, a fixed-$\mu$ four-lobe heat-capacity pattern, and opposite compressibility peaks at the two band edges.

The electrical boundary condition adds a second level of structure.  We formulated a constant-gate-voltage circuit, retained the full massive-Dirac density of states, and included the shift of the field-even heat capacity caused by $\mu_+\neq\mu_-$.  In the low-temperature fixed-number limit the edge value is $-2(\ln2)^2\DB\kb$, and a finite geometric capacitance interpolates smoothly to fixed $\mu$.  An independent finite-field pseudo-Landau-level sum, with the number constraint solved separately for the two orientations, agrees with the field-linear result at the representative controlled point.

A trace-free triaxial displacement supplies a concrete reversal protocol.  For a $156\,\mu$m-radius membrane, less than $0.74\%$ principal strain produces a $1.31$ mT uniform pseudofield while keeping $|\bs|A=100\ {\rm T}\,\mu{\rm m}^2$.  The same design places disorder, screened interaction, thermal, and circuit scales in a window where the single-particle crossover can be resolved.  Calorimetry and quantum capacitance then probe the same spectral flow through independent derivatives.  The framework applies to any multivalley Dirac material for which the species weights $\eta_a\zeta_a\sgn(m_a)$ add.

\begin{acknowledgments}
The research was supported by Armenian HESC grant 24FP-1F039.
\end{acknowledgments}

\appendix

\section{Pseudo-Landau levels and species signs}
\label{app:levels}

For one species, write
\begin{equation}
\begin{aligned}
H_a&=v_F(\eta_a\sigma_x\Pi_x+\sigma_y\Pi_y)+m_a\sigma_z,\\
[\Pi_x,\Pi_y]&=-i\hbar e\zeta_a\bs.
\end{aligned}
\label{eq:appH}
\end{equation}
The product $\eta_a\zeta_a\bs$ sets the oscillator orientation.  Solving the two-component Landau problem gives
\begin{equation}
E_{0a}=-\eta_a\zeta_a m_a\sgn(\bs),
\label{eq:appE0}
\end{equation}
and
\begin{equation}
E_{n,\pm,a}
=\pm\sqrt{m_a^2+2n\hbar e v_F^2|\bs|},
\qquad n\geq1.
\label{eq:appEn}
\end{equation}
Only $E_{0a}$ changes under field reversal.  The odd response of species $a$ is therefore proportional to $\eta_a\zeta_a\sgn(m_a)$.  This proves Eq.~\eqref{eq:selection}.

For graphene, $\eta_\tau=\tau$.  A real vector potential has $\zeta_\tau=1$, while a strain vector potential has $\zeta_\tau=\tau$.  With a Semenoff mass $m_\tau=m$, the real-field weights cancel and the strain-field weights add.  With a Haldane mass $m_\tau=\tau m_H$, the pattern is reversed.  This distinction is important when extending the result to other gap mechanisms.

\section{Thermodynamic derivatives and numerical checks}
\label{app:derivatives}

Starting from Eq.~\eqref{eq:freeenergy}, use
\begin{equation}
\frac{\partial x_\pm}{\partial\mu}
=\frac{1}{2\kb T},
\qquad
\frac{\partial x_\pm}{\partial T}
=-\frac{x_\pm}{T}.
\label{eq:xderivatives}
\end{equation}
A direct derivative gives Eq.~\eqref{eq:Cth}.  A second derivative with respect to $\mu$ gives Eq.~\eqref{eq:kappa}.  The entropy follows from $s=-\partial_T\omega$ and the identity
\begin{equation}
\frac{d}{dT}\left[T\ln\cosh x(T)\right]
=\ln\cosh x-x\tanh x.
\label{eq:entropyidentity}
\end{equation}
Differentiating once more gives the heat kernel $\phi=x^2\sech^2x$.  These steps also give Eq.~\eqref{eq:alpha}, and mixed derivatives prove Eq.~\eqref{eq:maxwell}.

\begin{figure}[t]
\includegraphics[width=\columnwidth]{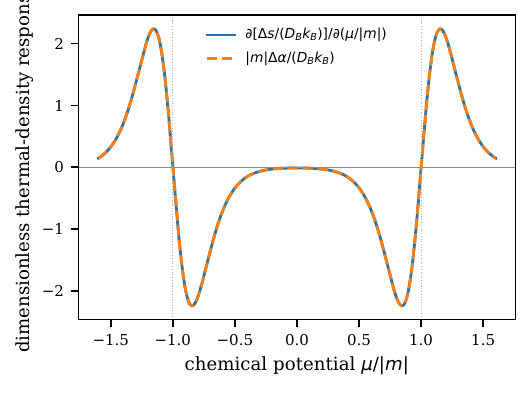}
\caption{Numerical check of the Maxwell relation.  A centered finite-difference derivative of the entropy agrees with the analytic thermal-density response.}
\label{fig:maxwell}
\end{figure}

Figure~\ref{fig:maxwell} compares the two sides numerically.  The same evaluation also verifies the derivative of $\cth$ against Eq.~\eqref{eq:kappa} and the extremum condition in Eq.~\eqref{eq:ystar}.

\section{Band-edge scaling functions}
\label{app:edgescaling}

At the conduction edge, $x_-=y$ and $x_+=M/(\kb T)+y$.  Since
\begin{equation}
g(x\to+\infty)=-\ln2,
\qquad
\phi(x\to+\infty)=4x^2e^{-2x}+\cdots,
\label{eq:asymptotics}
\end{equation}
Eqs.~\eqref{eq:edges}--\eqref{eq:edgek} follow.  The fixed-$\mu$ extrema satisfy
\begin{equation}
\frac{d}{dy}\left[y^2\sech^2y\right]
=2y\sech^2y(1-y\tanh y)=0,
\label{eq:peakderivative}
\end{equation}
which gives Eq.~\eqref{eq:ystar}.

\begin{figure}[t]
\includegraphics[width=\columnwidth]{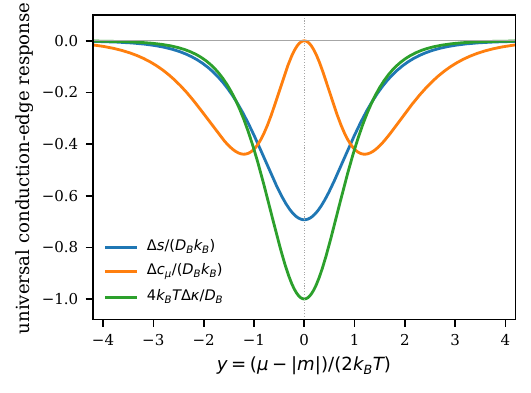}
\caption{Universal low-temperature scaling functions near the conduction edge at fixed $\mu$.  The entropy reaches $-\ln2$ at $y=0$, the heat capacity has an asymptotic node there, and the compressibility has its largest magnitude there.}
\label{fig:edgescaling}
\end{figure}

The exact finite-gap result at the nominal node is Eq.~\eqref{eq:finitegapnode}; it approaches the low-temperature asymptote exponentially as $M/(\kb T)$ grows.

\section{Gate thermodynamics and finite-field validation}
\label{app:ensemble}

For a graphene sheet coupled to a fixed-voltage gate, Eq.~\eqref{eq:gateconstraint} follows from the electrochemical relation
\begin{equation}
eV_g=\mu+\frac{e^2(n-n_g)}{C_g},
\label{eq:electrochemical}
\end{equation}
with all voltage-independent offsets absorbed into $\mu_{\rm res}$.  Differentiating at fixed $V_g$, $n_g$, and $C_g$ gives Eq.~\eqref{eq:dmudT}.  Along this path,
\begin{equation}
\frac{ds}{dT}
=\left(\frac{\partial s}{\partial T}\right)_\mu
+\left(\frac{\partial s}{\partial\mu}\right)_T
\frac{d\mu}{dT}.
\label{eq:dsdT}
\end{equation}
The Maxwell relation $(\partial s/\partial\mu)_T=\alpha$ then gives Eq.~\eqref{eq:cR}.  This derivation assumes that only the graphene electronic entropy follows the temperature modulation.  A thermally participating gate would contribute its own entropy and heat capacity, which are field even to leading order and must be included in the experimental background model.

For the zero-field massive Dirac cone, let $\theta=\kb T$ and $w(z)=f(z)[1-f(z)]$, with $f(z)=[e^{z/\theta}+1]^{-1}$.  Using Eq.~\eqref{eq:rho0}, the density relative to charge neutrality is
\begin{equation}
n_0=\int_M^\infty dE\,\rho(E)
\left[f(E-\mu)-f(E+\mu)\right].
\label{eq:fullDOSn}
\end{equation}
The even response functions used in Eq.~\eqref{eq:deltaCRgeneral} are
\begin{align}
\kappa_0={}&\frac{1}{\theta}\int_M^\infty dE\,\rho(E)
\left[w(E-\mu)+w(E+\mu)\right],
\label{eq:fullDOSkappa}\\
\alpha_0={}&\frac{1}{\kb T^2}\int_M^\infty dE\,\rho(E)
\Big[(E-\mu)w(E-\mu)\notag\\
&\hspace{24mm}-(E+\mu)w(E+\mu)\Big],
\label{eq:fullDOSalpha}\\
c_{\mu0}={}&\kb\int_M^\infty dE\,\rho(E)
\Bigg[\frac{(E-\mu)^2}{\theta^2}w(E-\mu)\notag\\
&\hspace{24mm}+\frac{(E+\mu)^2}{\theta^2}w(E+\mu)\Bigg].
\label{eq:fullDOSc}
\end{align}
Together with Eqs.~\eqref{eq:Cth}, \eqref{eq:cmu}, \eqref{eq:kappa}, and \eqref{eq:alpha}, these integrals produce Eq.~\eqref{eq:deltaCRgeneral} without a constant-density-of-states approximation.

For completeness, we give the low-temperature fixed-number reduction.  With a constant conduction-band density of states, define
\begin{equation}
p(y)=\frac{1}{1+e^{-2y}},
\qquad
L(y)=\ln(1+e^{2y}).
\label{eq:pLq}
\end{equation}
The background functions satisfy $\kappa_0=\rhoM p$ and $\alpha_0=\rhoM\kb[L-2yp]$.  Substituting them into Eq.~\eqref{eq:deltaCRgeneral}, including the final chemical-potential-shift term, gives Eq.~\eqref{eq:fixedn}.  At $y=0$ this reduces to $-2(\ln2)^2$.  The field-linear expansion is nonuniform deep in the gap because $\kappa_0$ becomes exponentially small; Eq.~\eqref{eq:weakflux} must always be checked.

The exact finite-field validation uses the full pseudo-Landau spectrum.  For $m>0$ and orientation $\sigma=\pm1$, $E_{0\sigma}=-\sigma M$ and $E_n=\sqrt{M^2+2n\hbar e v_F^2|\bs|}$.  The density relative to neutrality is
\begin{equation}
\begin{aligned}
n_\sigma(\mu) =\DB\Bigg[&f(E_{0\sigma}-\mu)-\frac12\\
&+\sum_{n=1}^{\infty}
\left(f(E_n-\mu)-f(E_n+\mu)\right)\Bigg].
\end{aligned}
\label{eq:LLdensity}
\end{equation}
The remaining response functions follow by analytic derivatives of the same occupations.  For a finite gate, the chemical potential is found from
\begin{equation}
\mu_\sigma-\mu_0
+\frac{n_\sigma(\mu_\sigma)-n_0(\mu_0)}{\Kg}=0.
\label{eq:exactgateconstraint}
\end{equation}
At fixed number one instead solves $n_\sigma(\mu_\sigma)=n_0(\mu_0)$; at fixed $\mu$, one sets $\mu_\sigma=\mu_0$.  Equation~\eqref{eq:cR} is then evaluated separately for $\sigma=+1$ and $-1$ before taking the difference.  This order of operations retains the nonzero pseudo-Landau levels and is the basis of Fig.~\ref{fig:ensemble}(c).

\bibliography{references}

\end{document}